\documentclass[twocolumn,showpacs,preprintnumbers,amsmath,amssymb,
aps,pra,10pt, 
]{revtex4-2} 

\usepackage{graphicx}
\usepackage{dcolumn}
\usepackage{bm}
\usepackage{color}
\usepackage{amsmath}
\usepackage{amssymb}
\usepackage{dsfont}
\usepackage{color}
\usepackage{calc}
\usepackage{hyperref}
\usepackage{xcolor}
\definecolor{AIPBlue}{RGB}{61, 180, 229} 
\hypersetup{
    colorlinks,
    linkcolor = AIPBlue,
    citecolor = AIPBlue,
    urlcolor = AIPBlue
}
\usepackage{multirow}
\usepackage{float}
\usepackage[normalem]{ulem}
\usepackage{natmove}

\def\be{\begin{equation}}
\def\ee{\end{equation}}
\def\bea{\begin{eqnarray}}
\def\eea{\end{eqnarray}}

\newcommand{\blue}{\color{blue}}

\graphicspath{{./figures/}}

\begin{document}

\title{Concept of introverted space: is extroverted, multidimensional space an illusion?}

\author{Ingo Steinbach}
 \email{ingo.steinbach@rub.de}
\affiliation{ Ruhr-University Bochum, ICAMS, Universitaetsstrasse 150, 44801 Bochum, Germany}



\date{\today}

\begin{abstract}
The quantum-phase-field concept of matter is revisited with special emphasis on the introverted view of space. Extroverted space surrounds physical objects, while introverted space lies in between physical objects. Space between objects leads to a network structure of matter: a network in which one-dimensional {\blue space filaments} connect massive elementary particles. {\blue Missing quantum fluctuations in the finite space filaments are interpreted as `gravitons,' the exchange particles of gravitational attraction between elementary particles.}
\end{abstract}

 \maketitle
{\blue This manuscript is revision of the paper `Concept of introverted space: is extroverted, multidimensional space an illusion?,' published in Zeitschrift fuer Naturforschung A https://doi.org/10.1515/zna-2023-0288.

The changes are marked in blue letters. One new aspect is the `Renormalization' procedure in chapter \ref{QPF}, which establishes discrete objects (particles) in the continuum theory of Quantum Phase Fields. Also the interpretation of missing quantum fluctuations in finite space as `gravitons' is added.}

\section{Introduction}\label{Intro}

`Space' is one of the first perceptions a child makes after being born: space separates the child from its mother. It is reported that babies first see the world upside down because of the optic our eyes. The image, generated by the brain from individual signals of the visual nerves, is understood as field in a two-dimensional (2D) vector space. Since we have two eyes, our brain can add the third dimension. These empirical observations over thousands of years have been formulated in the mathematical language of physics as the concept of fields in a continuous multidimensional vector space. There is no question about the practical success of this approach, which culminates in Newton's mechanics and Einstein's general relativity.

But is `space,' in which we are living, a multidimensional continuum on a fundamental level? Is it necessary to describe space as a {\blue multidimensional} vector space? What problems arise if we do so? What alternative descriptions are possible?

There are several approaches to the understanding general physics that are based on discrete spaces, string theory, loop gravity, or similar concepts. Typically, they embed discrete objects (objects with a reduced dimensionality) such as strings or branes into a high-dimensional continuous space-time. An infinite number of these discrete objects, so-called `quantum oscillators,' are attached to each region in multidimensional space.

My concept of a discrete space is different: it develops a network of masses and spaces that are formed by quantum phase fields. Herein, I shall:
\begin{itemize}
    \item {illustrate the introverted and extroverted views of space;}
    \item {define the network structure of masses and spaces;}
    \item {revisit the quantum-phase-field theory with emphasis to introverted spaces.}
\end{itemize}

\section{Introverted versus extroverted view of space}\label{IntroExtro}
Let us start with the extroverted view of space; this is the common view in traditional physics, including quantum physics and general relativity. Extroverted space is something that exists with or without massive particles---leptons and quarks---embedded into it. Within the extroverted view, particles are placed into space. In this view, space is described mathematically as a real vector space of $n$ dimensions, where $n \ge 3$. Empty space, i.e., space in which there are no particles, may not be seen as fully empty because there are quantum fluctuations. There are concepts that consider space to be related to some `substance,' called ether, and this may be related to these quantum fluctuations.

Space is generally seen to be filled by `fields': real or complex, scalar or vector valued, classical or quantum. These fields have a characteristic value at each point in space, and their best-known example is the electric field. Since extroverted space is formulated as a real and continuous vector space, classical fields have a measurable value `locally' at each point in this vector space. In quantum mechanics, one has to also consider the gradients of the fields, which makes the description `non-local.' In general relativity, one has to consider the invariance of the speed of light to connect three-dimensional (3D) real space with time, forming a 3D manifold in four-dimensional space-time. The fields transfer attractive or repulsive action between particles, and gravitation is not seen as an attractive force in general relativity, but as a consequence of the coupling between mass and the curvature of space-time.

\begin{figure}[ht]
 \centering
\includegraphics[width=8.6cm]{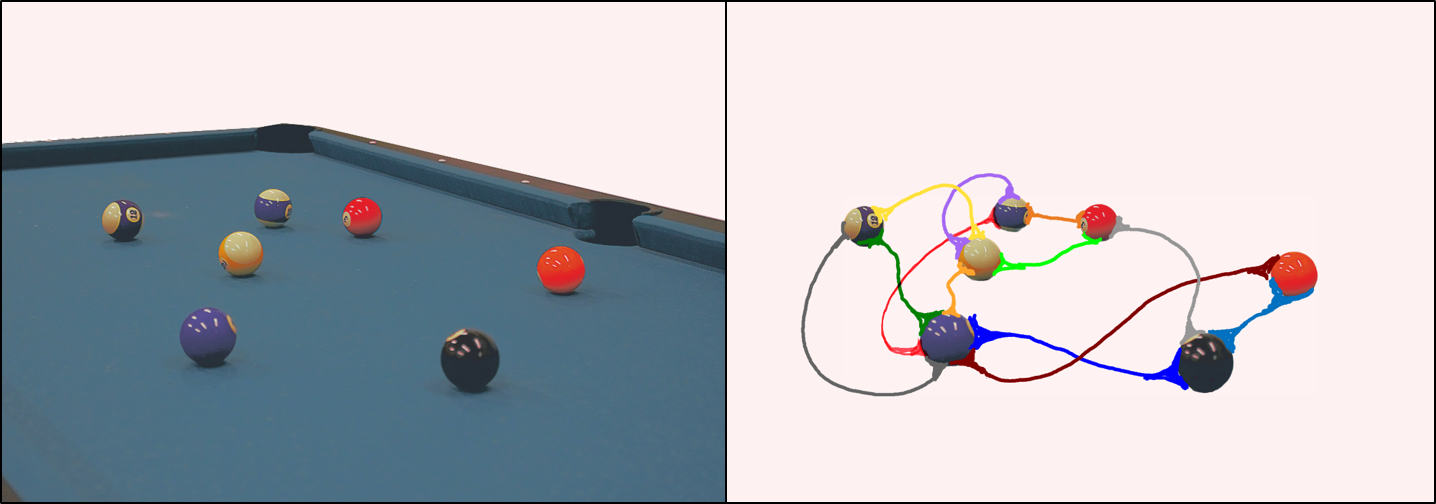}
\caption{Left: particles placed in space like billiard balls on a table. Right: particles and spaces as a network. Particles connect spaces (the line segments between the balls), and spaces connect particles. Note: The example on the left is 2D space (the table)
and 3D billiard balls with finite volume; This is meant to be generalized to a network with point masses (nodes) and the coloured 1D spaces (edges), right figure.}\label{Billiard}
\end{figure}

I compare this to a billiard game (although in this analogy, action is only transferred when balls collide, but the indentation of the ball into the cloth may be of importance). The green of the billiard table represents space. The particles---billiard balls---are placed in this space (see Figure~\ref{Billiard}, left). They interact according to Newton's laws of momentum and energy conservation, or their relativistic extensions. In a real billiard game, the player uses the cushions to mirror space. We may, hypothetically, push the cushions to infinity; then, our playground will be infinitely large. We may use periodic boundary conditions, or we might generalize flat space to Riemann geometries of different topology. The basic principle of the extroverted view of space stays the same: space exists with or without particles. Particles are placed into space, and space `surrounds' these particles. I thus call this the extroverted view.

The introverted view is different; here, space, if you will, separates and connects particles (Figure~\ref{Billiard}, right). I call this view `introverted' because it considers that space lies \textit{between} particles. The space inside a building, which is mostly called `room' in this context, is surrounded by walls and floors. Walls and floors are 2D objects in a 3D world; these objects do not exist as elementary entities in physics. Particles---leptons or quarks---are zero-dimensional, meaning  they are so small that no extension in any direction can be attributed to them, they are point like. Therefore, introverted space lying in-between the particles has to be one-dimensional (1D): there is no other choice. Particles may then be represented as the vertices, or nodes, of a network structure. The connections---the edges of the network---can be seen as `spaces,' and they are each defined by the distance between the two particles that they connect.

Here, we must take a moment to reflect: what does `distance' mean? It is first of all a scalar value that is assigned to the relation between exactly two objects. The distance is small if the objects are close to each other, and it is large if they are not close. Considering `not close' in contrast to `close,' let us say that they have less to do with each other, that they interact less. We may measure the distance by a length in $[\mathrm{m}]$ or by a time to transfer action in $[\mathrm{s}]$. We may also measure it according to the strength of interaction: by the binding energy between two particles in $[\mathrm{J}]$. We will {\blue in the following} understand the interrelation of two particles being related to an energy! \footnote {Particles are usually associated by a positive energy of their rest mass. Distance, space between particles, commonly, is not considered as an energetic state, although Newtons physics clearly draws this connection. I show that space has a negative energy, the energy of space, see section \ref{energy_density}} 

The quantum-phase-field concept, as reviewed in this essay, describes mass and space as a network of energetic states. It is published in \cite{Steinbach_QPF,Steinbach_Soliton}, and a common version can be found in Chapter~8 of \cite{Lectures}.

\section{The network of masses and spaces} \label{DoublonNetwork}

Let us start with a formal definition of the network of particles and their connecting spaces.

The network is defined by nodes (particles) $U_i$, $i = 1 \ldots n$, and edges (spaces) $E_I$, $I=1 \ldots N$. For a fully connected network, we have $N=\binom{n}{2}$, but we will also consider partially connected networks $N<\binom{n}{2}$. We postulate the following topology as the simplest form consistent with the above principles:
\begin{itemize}
\item One $E$-element $E_I \Doteq E_{km}$ connects two U-elements $U_k$ and $U_m$, $k, m \in 1 \ldots n$.
\item One $U$-element $U_i$ connects a number of ${\blue 4} \le M \le N$ $E$-elements $E_K$, $K \in 1 \ldots N$.
\end{itemize}

Clearly, this leads to a network as depicted in Figure~\ref{Billiard}, right, where the $U$-elements are nodes and the $E$-elements are edges. Note that a minimum of {\blue four nodes and six edges is needed} to form a primitive network {\blue in three dimensions. We will, without loss of generality, later also treat the case of} a ring of two edges connected by two nodes, which can be solved analytically. {\blue I} will, however, mostly speak about \textit{many} nodes and edges. Without loss of generality, we associate the $U$-elements with positive energy, the massive energy of particles, and the $E$-elements with negative energy, the energy of space. 
In the next chapter I will relate `edges' and `nodes' to `quantum phase fields'.

\section{Formal definition of quantum phase fields} \label{QPF}
The basic object in the quantum-phase-field concept is a `phase.' The phase distinguishes a piece or matter from other pieces in a different phase state; i.e., there is at least one attribute of the piece of matter in one particular phase state that distinguishes it from other pieces in a different phase state.

In quantum physics we call the
phase a ‘quantum state’ and each state can be occupied only once. In condensed matter physics, the phase characterizes atomic order, or its absence: solid, liquid, gas, plasma, etc.; this also applies to magnetic order and so on. Each phase is distinguished from another phase by a so-called `order parameter,' which is normalized to $0\le \phi \le 1$.

We will allow \textit{many} phases, $\phi_I$, $I=1 \ldots N$, and each phase is connected to all other phases $\phi_K$, $K\ne I$:
\be\label{universe}
\sum_I \phi_I = 1.
\ee
This is a system without outer boundaries; the phases $\phi_I$ constitute a closed system, forming a `universe'. We do this in analogy to the multi-phase-field theory in condensed matter physics \cite{Steinbach99b}.

Here, $\phi_I$ is simply termed `phase' because so far, no space has been defined. Each phase $\phi_I$ is an element of the system that is different from all other elements $\phi_K$, but there are connections between the phases in situations where at least two phases have a value $\phi_I < 1$ due to the constraint \eqref{universe}. In fact, we will identify the pure phases $\phi_I = 1$ with the edges of the network $E_I$.

The task now is to define the nodes $U_i$. If the phase forms an edge, it must connect exactly two nodes. However, one node may connect many edges. Later, when `space' is introduced, we will simply denote all points in space where at least two phases are connected (with the condition $\phi_I < 1$, ${\blue \sum_{k=1, k \ne I}^N} \phi_k = 1 - \phi_I$) as `nodes.'

To give the phases a physical meaning, I associate them with the conserved quantity `energy' $H = \langle \psi| \hat H |\psi\rangle$. Here, $|\psi\rangle$ is the quantum state of the system and $\hat H$ is the energy operator. Furthermore, I postulate that $H=0$, i.e., there is no net energy: all energetic states, positive and negative, have to sum to zero. The argument for this is simple: there is no evidence regarding where a finite energy of the universe should come from (see also the Wheeler--DeWitt theory \cite{DeWitt1967}).

I allow changes $\mathrm{d} \hat H$ such the zero-energy state---the state of `nothing'---separates into positive and negative energetic elements---the state of `something.' This can be related to the Big Bang as the origin of our universe, if you will. {\blue We expand $\hat H$ in the changes $\mathrm{d} \hat H$ with respect to the phases $\phi_I$. $\hat H$ is thus itself a function of all phases $\{\phi_I\}$, $\hat H = \hat H(\{\phi_I\})$, and, as a reminder, all fields are connected by the sum constraint \eqref{universe}. We start with the identities:
\begin{subequations}
\bea\label{Expansion_H}
\hat H = \sum_{I=1}^N \hat H_I \\
\hat H_I = \int_0^1 d\phi_I \frac{\partial \hat H_I}{\partial \phi_I}. \label{2_3}
\eea
\end{subequations}

The integral runs over the definition range of the phase as an order parameter from $0$ to $1$, meaning \textbf{yes} or \textbf{no}, \textbf{existing} or \textbf{not-existing}. Here, again, we need to pause for a moment: Relation \eqref{2_3} simply means that $H_I = H_I(\phi_I=1)$ if $\phi_I$ is seen as a scalar variable (we postulate for a non existing phase $\hat H_I(\phi_I=0) = 0$).  In order to consider changes in the phase structure, we allow the phases to vary between their bounds; Then, we have to introduce another variable which, as the simplest case, shall be a scalar variable. And we treat the phase a function of this variable, thus becoming a field with this variable as a coordinate. The phase becomes a phase field and the variable will be considered as a length coordinate in the following.}

Introducing the length coordinate $s_I$ (corresponding to the size of an edge of the network), substituting $\mathrm{d}\phi_I = \dfrac{\partial\phi_I}{\partial s_I} \mathrm{d}s_I$, and introducing the forces {\blue $\hat h_I = \dfrac{\partial \hat H_I}{\partial s_I}$} yields
\bea\label{energy_op}
\nonumber \hat H &=& \sum_{I=1}^N \int_{-\infty}^\infty \mathrm{d}s_I \frac {\partial \phi_I}{\partial s_I} \frac {\partial \hat H_I}{\partial \phi_I}\\
&=& \sum_{I=1}^N \int_{-\infty}^\infty \mathrm{d}s_I \hat h_I.
\eea

Space emerges; i.e., it is created by variations in the phases $\mathrm{d} \phi_I$. We relate the line coordinate $s_I$ to the distance $\Omega_I$, which is defined by the negative inverse of the energy of an edge $E_I$ (for the self-consistent proof see Section~\ref{energy_density}) by the integral
\be\label{s_Omega}
\int_{-\infty}^{+\infty} \mathrm{d}s_I \phi_I(s_I) = \Omega_I = - \tilde \alpha \frac{hc}{48 E_I},
\ee
where $\tilde \alpha$ is a dimensionless parameter to be defined, $h$ is Planck's quantum and $c$ speed of light.

It is important to note that the space $s_I$ is not `fundamental,' but it is defined by the fundamental entity `energy' as a real number {\blue with unit [m]}. It is an auxiliary coordinate to link the concept to the view of physics rooted in wave mechanics. From now on, we will consider the phase $\phi_I$ as a field $\phi_I(s_I)$ in the line coordinate $s_I$ that is intrinsic to this field.

The force operator $\hat h$ is expanded in the phases and in their gradients:
 \bea
\nonumber \hat h_I = u &\Bigm ( & \eta^2 \left[ \left( \frac \partial{\partial s_I} \phi_I\right)^2 - \frac1{c^2}\left(\frac \partial{\partial t} \phi_I\right)^2\right] \\
\label{force_op}
 &+& P_I(\{\Phi_J\}\Bigm ),
\eea
\be\label{Potential}
{\blue P}_I = \left (\gamma \sum_{\substack{J=1, \\ J\ne I}}^N\phi_I^2 \phi_J^2 + \tilde \gamma \sum_{\substack{J,K,l=1, \\ I\ne J \ne K \ne L}} ^N\phi_I \phi_J\phi_k\phi_l \right).
\ee

Here, $u$ is a positive constant with dimensions of energy per length, or force $\left[\frac{\mathrm{J}}{\mathrm{m}}\right]$, and $\eta$ is a positive length quantum. The gradient contribution in time is included, which ensures relativistic invariance due to Lorentz contraction of the length $\eta$ (see \cite{Steinbach_QPF} for details). Furthermore, ${\blue P}_I(\{\Phi_J\})$ is the Landau potential for phase $I$ expanded in all connected phases $J$. In contrast to previous publications \cite{Steinbach_QPF,Steinbach_Soliton}, we employ a $\phi^4$ potential in the current version of the theory, with force parameters $\gamma$ and $\tilde \gamma$, for discussion see \cite{Steinbach2020}.

The second potential term connecting the four different phases in Eq.~\eqref{Potential} has an intriguing consequence: this is the only surviving potential term in the limit $N \rightarrow \infty$. It states that there is a minimum of four phases to be connected, with $\phi_I\ne\phi_J\ne\phi_K\ne\phi_L$, which forms the minimum requirement for a space-filling body in three dimensions! All terms with lower numbers of connections vanish in the limit $N \rightarrow \infty$. To see this, we investigate the center of a node at which all connecting phases are equal $\phi_I=\phi_J = \dfrac 1 N$. In equilibrium, this center has the highest energy. The first term in the expansion of the Landau potential \eqref{Potential} is the sum over pairs $\phi_I^2 \phi_J^2 \approx \dfrac 1{N^4}$. Since there are $\dbinom{N}{2}$ contributions, and $\dbinom{N}{2}\rightarrow \dfrac 12 N^2$ for $N\rightarrow \infty$, this term vanishes for large networks. I retain this here for the analytical solvability of one edge between two nodes. The second term in the potential \eqref{Potential}, $\phi_I \phi_J\phi_k\phi_l$, remains of order unity since $\dbinom {N}{4} \rightarrow \dfrac 1{24} N^4$ for $N \rightarrow \infty$. Higher-order terms, $\phi^6$, $\phi^8$, and so on, may be added with the same argument, as long as all phases are independent and we sum over all combinations.\footnote{Using the same argument, in previous versions of the theory \cite{Steinbach_QPF,Steinbach_Soliton}, we restricted ourselves to a pair of phases and a quadratic potential. Here, however, a non-analytical treatment of the non-linearity of the phase-field equation is needed, see \cite{Steinbach2020}.} 

The gradient contributions of the energy operator \eqref{force_op}, $\dfrac \partial{\partial s}$ and $\dfrac 1c \dfrac \partial{\partial t}$, shall be understood as operators acting either on the phase fields or on the quantum mechanical wave function $|\psi\rangle$. The kinetic equation for the evolution of phase fields is written down according to the Clausius--Duhem relation, with the relaxation constant $\tilde \tau$:
{\blue
\be\label{statistical_motion}
 \tilde \tau \frac\partial{\partial t} \phi_I = - \frac{\delta}{\delta\phi_I}   \int_{0}^{+\infty}dt \left [ \sum_{J=1}^N \int_{-\infty}^{+\infty}ds_J \langle \psi| \hat h_J|\psi\rangle \right ].
\ee }

This equation has two parts: (i)~a non-linear wave equation for the phase fields $\phi_I$, and (ii)~a linear Schr\"{o}dinger-type equation for quantum-mechanical excitations {\blue $\psi$ }. This procedure is not new; it can be traced back to the so-called de~Broglie--Bohm double-solution program \cite{Bohm1952a,Bohm1952b,deBroglie1960,deBroglie1971}. For further explanation, see \cite{Steinbach_Soliton}.

We now separate the expectation value of the energy operator (\ref{energy_op}) into three different contributions. These are distinguished by whether the differential operators $\frac {\partial}{\partial s}$ and $\frac {\partial}{\partial t}$ are applied to the wave function $|\psi\rangle$ or the field $\phi_I$.

Applying the differential operators to the phase components and using the normalization of the wave function $\langle \psi|\psi \rangle=1$ yields the force $u_I$ $[\frac{\mathrm{J}}{\mathrm{m}}]$ related to phase $I$ {\blue, with $ p_I = \frac \partial{\partial \phi_I} P_I$.}:
\be
\label{classical}
 u_I = u \left\{ \eta^2\left[\left(\frac \partial{\partial s_I} \phi_I\right)^2 - \frac1{c^2}\left(\frac \partial{\partial t} \phi_I\right)^2\right] + p_I \right\}.
\ee



The mixed contribution, in which one of the operators $\frac{\partial}{\partial s}$ and $\frac{\partial}{\partial t}$ is applied to the field $\phi$ and one is applied to the wave function $|\psi\rangle$, describes the correlation between the field and the wave function. This shall be set to $0$ in the quasi-static limit. In this limit, we keep the field static for the evaluation of the quantum-mechanical force. Then, we take this force for the determination of the time evolution of the field. A coupled solution has not been worked out to date:
\be\label{acceleration}
0= \phi_I u\eta^2 \left[ \frac {\partial \phi_I}{\partial s_{\blue I}} \langle \psi|\frac {\partial}{\partial s_{\blue I}}|\psi\rangle - \frac1{c^2}\frac {\partial \phi_I }{\partial t}\langle \psi|\frac {\partial}{\partial t}|\psi\rangle \right].
\ee
It is shown in \cite{Steinbach_QPF} that Eq.~\eqref{acceleration} is consistent with Newton's second law of acceleration. Finally, we apply the momentum operators $\frac{\partial}{\partial s}$ and $\frac 1c \frac{\partial}{\partial t}$ to the wave function $|\psi\rangle$, which yields the force $e_I$ $\left[\frac{\mathrm{J}}{\mathrm{m}}\right]$:
\be\label{energydensity}
e_I =u\eta^2 \phi_I^2 \langle \psi|\frac {\partial^2}{\partial s_{\blue I}^2}-\frac1{c^2}\frac {\partial^2}{\partial t^2}|\psi\rangle.
\ee

This contribution applies to the bulk energy of the phase field $\phi_I= 1$. We will explicitly evaluate this after the structure of the solutions of the fields is discussed. One transforms the phase-field equation \eqref{statistical_motion} into the moving frame traveling with the velocity $v$. Inserting Eqs.~\eqref{classical}--\eqref{energydensity} into \eqref{statistical_motion}, we find:
\bea\label{motion_MF}
\nonumber \tilde \tau \frac\partial{\partial t} \phi_I &=& -\frac{\delta}{\delta\phi_I} \int_{0}^{+\infty}\mathrm{d}t \langle \psi| \hat H|\psi\rangle\\
\nonumber &=&
u \left[\eta^2 \frac {\partial^2\phi_I}{\partial s_{\blue I}^2}\left(1-\frac{{v}^2}{c^2}\right) - {\blue p_I}\right] \\
\label{TW} &+& m_{\phi_I}\sum_{j=1}^N\Delta e_{IJ},
\eea
where: $\Delta e_{IJ} = e_I - e_J$ is the difference in the volume force between two phases $I$ and $J$ according to Eq.~\eqref{energydensity}; and $m_{\phi_I}$ is the appropriate {\blue dimensionless} coupling function.

\begin{figure}[ht]
\centering
\includegraphics[scale=0.3]{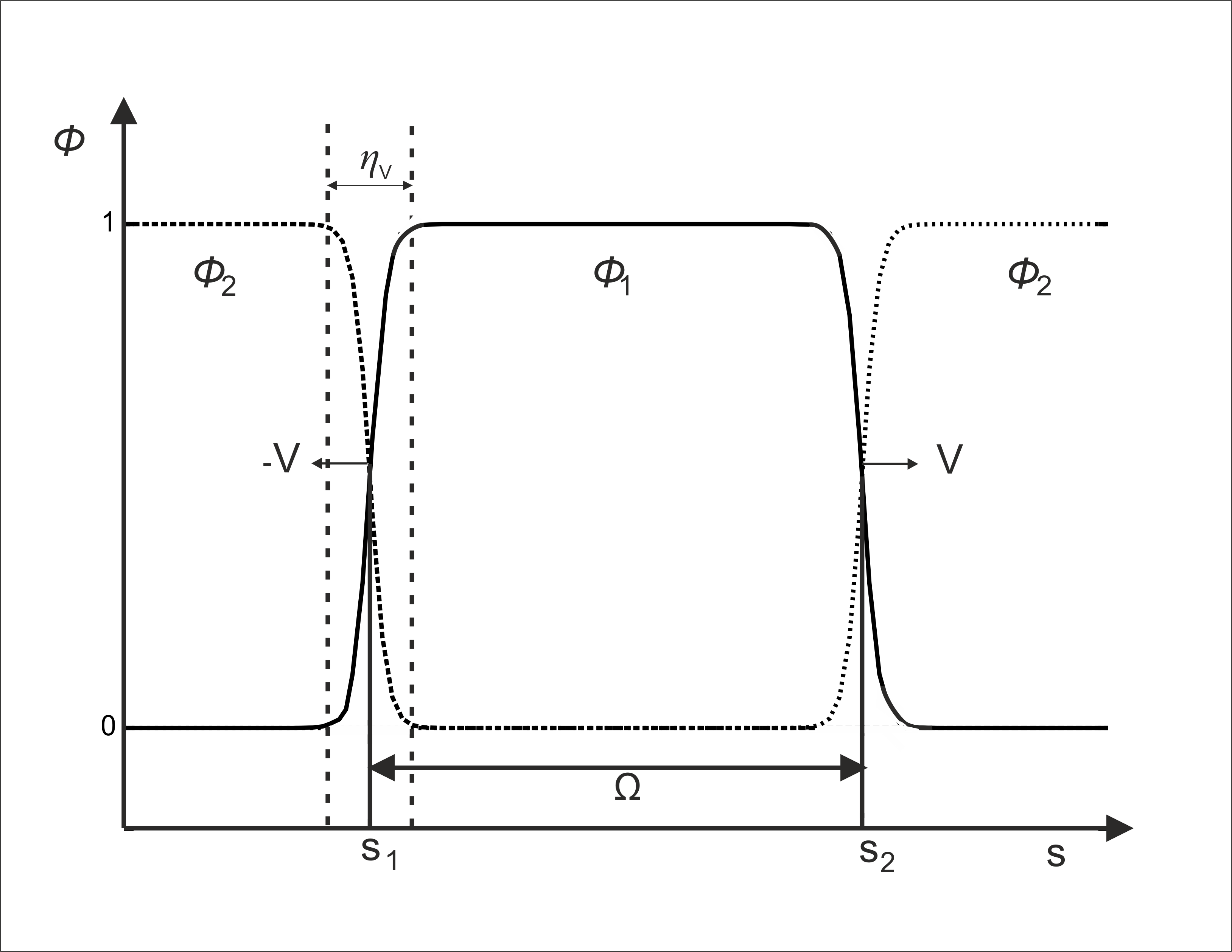}
\caption{Two doublons in a periodic setting. Each doublon is formed by a right-moving and a left-moving soliton. The velocity is proportional to the energy difference between adjacent doublons. The relative distance $\Omega_{12} = s_1-s_2$ corresponds the length of the edges of the network. }
\label{fig:doublon}
\end{figure}

The so-called doublon solution for two phases in a periodic setting is well known as the minimum solution of the classical part of the energy operator $\hat H$ [Eq.~\eqref{energy_op}], as depicted in Figure~\ref{fig:doublon}:
\begin{subequations}
\be\label{extremal} 
{\blue
\eta^2 \frac {\partial^2\phi_I}{\partial s_{\blue I}^2}\left(1-\frac{{v}^2}{c^2}\right) - \phi_I(1-\phi_I)(\frac 12 - \phi_I) = 0 }
\ee
\bea
\label{travellingWave2}
\nonumber \phi_I(s_{\blue I}) &=&\frac12 \left\{ \tanh \left( \frac{3 \left(s_{\blue I} - s_1 -v t\right)}{ \eta_v}\right) \right.\\
&-& \left.\tanh \left(\frac{3 \left(s_{\blue I} - s_2 + v t\right)}{\eta_v}\right)\right\},
\eea
\end{subequations}
where the particles are located at $s_1$ and $s_2$ with distance $\Omega_{12} = s_1-s_2$. The phases transform into each other with velocity $v~\propto~\Delta~e$. $\eta_v = \eta\sqrt{1-\frac{v^2}{c^2}}$ shows Lorentz contraction of the quantum length for accelerated particles (for details, see \cite{Steinbach_QPF}). In this picture, $u$ is the force of inertia. One phase is bounded by two solitonian waves, one right-moving and one left-moving. We call this object a doublon. {\blue

\subsection*{Renormalization}\label{renormalization}
The final task is to renormalize the model parameter in a way that the theory becomes, within certain limits, independent of the size of the transition region $\eta$ between phase fields. We do this only for the low velocity case $v \rightarrow 0$, or for nodes in their resting frame. The renormalization procedure in the relativistic case $v \rightarrow c$ yet has to be worked out.

The energy quantum $U = u\eta$ shall be the principal entity of the system. It corresponds to the rest mass of the particles and shall be independent of particle size. Correspondingly we define the potential function $\tilde P_I = \frac {P_I} {\eta^2}$ and the relaxation parameter $\Theta = \frac {\tilde \tau} \eta$ as a constant. This corresponds to the physical notion, that the original relaxation time $\tilde \tau$ is proportional to the size of the object. The renormalization procedure defines discrete objects in a continuum theory.

The equation of motion \eqref{motion_MF} then reads after renormalization: 

\begin{subequations}
\bea\label{motion_MFr}
\nonumber \Theta \frac\partial{\partial t} \phi_I &=& 
U \left[ \frac {\partial^2\phi_I}{\partial s_{\blue I}^2}\left( 1-\frac{{v}^2}{c^2} \right) -  \tilde P_I \right] \\
\label{TW} &+& M_{\phi_I}\sum_{j=1}^N\Delta e_{IJ},
\eea
with an appropriate coupling functions $M_{\phi_I} = \frac {m_{\phi_I}}\eta $: 

\be\label{coupling}
M_{\phi_I} = \frac 6 \eta \phi_I(1-\phi_I) = \left | \frac {\partial \phi_I} {\partial s_I} \right |,
\ee 
\end{subequations}
where the latter identity can be read from the solution \eqref{travellingWave2} at one node point.}

\section{Volume energy of one doublon} \label{energy_density}
From the doublon solution in Section~\ref{QPF}, we can see that the field forms a 1D box with fixed walls and size $\Omega_I$ for phase $I$. We consider massless quantum fluctuations inside the box and find the explicit representation of the standing waves $\psi_p$ in the length coordinate $s$,
\be\label{psi}
\psi_p = \sqrt{\frac 2 \Omega} \sin\left({ \frac{\pi p}\Omega s}\right).
\ee

The dispersion relation for massless particles is linear in momentum $\tilde p= p \dfrac h \Omega$, $p \in \mathbb{N}$, instead of quadratic in $\tilde p$, as for the case of massive particles in a box potential. According to Casimir \cite{Casimir1948}, we have to compare quantum fluctuations in the box with discrete spectrum $p$ and frequency $\omega_p= \dfrac {\pi c p }{2\Omega_I}$ to a continuous spectrum. This yields the negative energy $E_I$ of the space $I$ {\blue as the integral over the energy density $e_I = \frac {\partial E_I}{\partial \Omega_I}$}:
\be\label{EC}
E_I = \alpha \frac {h c}{4\Omega_I} \left[ \sum_{p=1}^\infty p - \int_1^{\infty} p \mathrm{d}p \right] = - \alpha \frac {h c}{48\Omega_I},
\ee
where $\alpha$ is a positive, dimensionless coupling coefficient. I have used the Euler--Maclaurin formula in the limit $\epsilon \rightarrow 0$ after renormalization $p \rightarrow p e^{-\epsilon p}$. Since all parameters are positive, we see that `space' is accounted for by negative energy, scaling in inverse proportion to the size of the doublon, which proves \eqref{s_Omega} for self-consistency. This energy scales like the energy of the gravitational field in Newtonian mechanics and general relativity. Therefore, the force---as the derivative of the energy \eqref{EC} with respect to space---can be associated with a gravitational attraction between the nodes, which is transmitted by the spaces. {\blue In general, we may assotiatge the `missing' quantum fluctuations with massless bosonic excitation, called `gravions' from now on. Gravitons have negative energy, which can be understood in a similar way as the holes in the conduction band of a semiconductor being identified with positiv charges. Gravitons transfer the attractive gravitational force}. 

The nodes can thus be interpreted as massive objects: elementary particles {\blue which are attracted my gravitational forces (except on ultra short and ultra long distances, see below)}. They are associated with positive energy $U$. In contrast to Newtonian mechanics and in agreement with general relativity (see `gravitational waves' \cite{Einstein1918}), the attraction is a wave phenomenon with time-dependent action. The coefficient $\alpha$ can be determined from the measured gravitational constant on Earth, as we will derive in the next section.

\section{Closed doublon network}\label{doublon_network}
From the doublon solution \eqref{travellingWave2}, we see that the elements of the network---particles with energy $U_i$ and spaces with energy $E_I\equiv E_{ij}$---are both related to the phases, or if we relate the bulk energy of a phase to a space coordinate, we say that they are related to phase fields. Particles relate to the gradient contributions, while spaces relate to the bulk in between particles. Values of $\phi \equiv 0$ have no physical meaning. Since each doublon decays to zero on either side, it has to connect to at least one other doublon on each side due to the sum constraint \eqref{universe} within a finite transition region of size $\eta_v$. To get a handle on this size, as well as on the size of the bulk region $\Omega_{ij}$, we need to evaluate the coupling constant $\alpha$. We assume isotropy in the system and energy neutrality, with $N$ being the number of particles connected to each single particle, dropping the subscript $i$. Since only $\frac 12$ of each doublon counts for one particle, we find:
\be\label{neutrality}
0 = U + \sum_{I=1}^{N}E_{I} = U_i - \frac 12 \alpha \frac{hc}{48} \sum_{I=1}^{N}\frac 1 {\Omega_{I}}.
\ee
With the mass of one particle $m = \frac{U}{c^2}$ and the average size of the network as the homogeneous mean $\bar \Omega = N \left [\sum_{I=1}^{N}\frac 1 {\Omega_{I}}\right]^{-1}$, we get:
\be\label{alpha}
\alpha = 96 \frac{m c \bar\Omega}{N h}.
\ee
The force between two particles along one connecting doublon is evaluated:
\bea\label{microforce}
\nonumber
f_{I}^{\mathrm{micro}} &=& -\left(\frac{\partial E_{I}}{\partial\Omega_{I}}+ \frac{\partial E_{I}}{\partial\bar\Omega}\frac {\partial\bar\Omega}{\partial\Omega_{I}} \right) \\
\nonumber &=& \frac{ m c^2 \bar\Omega}{ N}\frac {1}{\Omega_{I}^2}\left[ 1-\frac{\bar\Omega}{ N \Omega_{I}} \right] \\
&=& \frac{ m^2 c^2 \bar\Omega}{ M^u}\frac {1}{\Omega_{I}^2}\left[ 1-\frac{\bar\Omega}{ N \Omega_{I}} \right].
\eea

The number of particles $N$ can be estimated from the mass in the visible universe $M^u$ compared to the mass of one particle $m$. For distances $\Omega_{I} < \dfrac{\bar\Omega}{N}$, we note repulsive action with a force scaling with $\dfrac 1 {\Omega_I^3}$. Thus, the singularity of collapse of an agglomeration of point-like particles into one point is forbidden. For $\Omega_{I} \gg \dfrac{\bar\Omega}{N}$, Newton's law of gravity is recovered. Identifying the prefactor in Eq.~\eqref{microforce} $\dfrac{c^2 \bar\Omega}{ M^u}$ with the coefficient of gravitation $G$, one obtains:
\bea\label{G-via-Omegabar}
\bar\Omega&=&\dfrac{G M^u}{c^2} \approx 7.4 \times10^{24}\, {\rm m} \approx 240\, {\rm Mpc}, \\
\dfrac {\bar\Omega} N &=& \dfrac{G m}{c^2} \;\; \approx 3 \times10^{-53}\, {\rm m}.
\eea

I have used the numerical values $G \approx 6.67\times 10^{-11} \left [\frac {\mathrm{m}^3}{\mathrm{kg} \cdot \mathrm{s}^2} \right ]$, $c \approx 3 \times 10^{8} \left [\frac {\mathrm{m}}{\mathrm{s}} \right ]$, $M^u = 10^{52} \left[\mathrm{kg}\right]$ \cite{Persinger2009}, and $m = 4\times 10^{-26} \left [\mathrm{kg} \right ]$ set to $\frac 14$, the mass of a hydrogen atom, or a neutron and its neutrino, each consisting of four fermions. Repulsive gravitational action at the microscale is thus limited to distances $\dfrac {\bar\Omega}N$ below the Planck length. We have, however, to consider that this formal derivation considers the variation of a cosmological length $\bar \Omega$ with a very short length scale, e.g., between quarks inside a nucleon. From the formal definition of $\bar \Omega$ by the harmonic mean of all spaces $\Omega(I)$, which is dominated by short spaces, we see that $\bar \Omega \rightarrow 0$ for one single $ \Omega(I) \rightarrow 0$, regardless of the large number of long and ultra-long spaces. The cutoff $\dfrac {\bar\Omega}N$ prevents this behavior, but on an unrealistically small scale, according to physical intuition. Therefore, the formal derivation of the micro force equation \eqref{microforce} has not been presented in previous publications \cite{Steinbach_QPF,Steinbach_Soliton}.

At `cosmological' distances, we shall treat $\bar\Omega$ and $\Omega_{I}$ as independent. Variation of the energy of space \eqref{EC} with respect to $\bar\Omega$ and $\Omega_{I}$ independently gives:
\bea\label{ForceClosed}
\nonumber f_{I}^{\mathrm{macro}} &=&-\left(\frac{\partial E_{I}}{\partial\Omega_{I}}+ \frac{\partial E_{I}}{\partial\bar\Omega}\right) \\
\nonumber &=&\frac{m c^2 \bar\Omega}{ N\Omega_{I}^2}\left[1-\frac{\Omega_{I}}{\bar\Omega}\right]\\
&=&G \frac {m^2}{\Omega_{I}^2}\left[1-\frac{\Omega_{I}}{\bar\Omega}\right].
\eea
Structures beyond the marginal distance $\bar\Omega$ repel each other, leading to an accelerating expansion. We further see from the generalized gravitational law \eqref{ForceClosed} that in the limit $\Omega_{I} \gg \bar\Omega$, the force scales as $f_{I} \propto - \dfrac 1 {\Omega_{I}}$ instead of $f_{I} \propto \dfrac 1 {(\Omega_{I})^2}$ for medium distances; i.e., repulsive gravitational action at ultra-long distances decays more slowly with distance than attractive gravitational action at medium distances. The consequences of this statement deserve further consideration in the future.

\section{Conclusion and discussion}\label{discussion}
Phases $\phi_I$ and their variations $\mathrm{d} \phi_I$ are the principal elements of the quantum-phase-field concept, forming a monistic theory. The model defines a network of particles and spaces that are both determined by the phases and their variation. Particles are not \emph{placed in} space, as in traditional theories, but rather they \emph{bound} spaces! Spaces connect particles, as particles connect spaces. Particles and spaces are two sides of the same coin: the doublon. One space element connects exactly two particles; one particle is the connection between many spaces. This defines the network structure of the physical world. Both elements of the network---nodes and edges---are defined by energy, which is the only fundamental substance in the concept. Space is introduced as a negative inverse of the energy of one edge. It is shown that this ansatz leads to a quantum problem on the edges, the solution of which reproduces the negative energy associated with this space self-consistently. The quantum problem is defined as a linear Schr\"odinger-type equation. The equation of motion for the nodes is formulated as a classical non-linear wave equation derived from a Ginzburg--Landau-type Hamiltonian on the line coordinate of space. The non-linearity suppresses the spectrum of solutions, as is characteristic for linear Schr\"odinger-type wave solutions. This can also be compared with the theory of solitonian waves \cite{Willox} and Goldstone modes in elementary particle physics \cite{Goldstone1961,Goldstone1962}.

The doublons are the minimum solution of the classical part of the quantum-phase-field equation \eqref{motion_MF}. The quantum part \eqref{energydensity} defines gravitational attraction or repulsion; the mixed part defines the equation of motion of particles under gravitational action.

Energy is the only fundamental substance, and its net amount is zero, balancing the positive energy of particles with the negative energy of space. The problem then arises that a closed quantum system without energy has to be stationary according to the time-dependent Schr\"odinger equation (cf. the Wheeler DeWitt equation \cite{DeWitt1967}):
\be\label{td_Schroedinger}
i h \dfrac{\partial \psi }{\partial t} = \hat H \psi = 0.
\ee
In the present concept, the problem is separated into a linear quantum problem on the individual doublons and a non-linear classical wave problem for the solitonian fronts. 
Equation~\eqref{td_Schroedinger} is linear in $\psi$, and time is seen as the observable conjugate to energy in the quantum-mechanical sense, and this part of the equation is reversible in time $t$.

$\hat H$ itself is a non-linear function of the classical field variables $\phi=\phi(\tilde t)$, and $\tilde t$ shall be called `thermodynamic time' as the observable conjugate to entropy. This distinction is introduced here to emphasize the different meaning of time in quantum mechanics (reversible in time) and thermodynamics, where time $\tilde t$ is not reversible according to the second law of thermodynamics. The thermodynamic time governs the dynamic evolution of the system according to the non-linear classical wave equation \eqref{statistical_motion} or \eqref{TW}. The evolution of the phases $\phi$ then changes the spectrum of quantum fluctuations within the doublons; i.e., it determines the time dependence of the wave function $\psi$. Although up until now only the quasi-static solution has been worked out, a coupled solution should exist. We may formally introduce a complex time variable, where $\tilde t$ forms the real axis and $i\,t$ the imaginary axis (see also the considerations in \cite{Chiatti}). I leave further discussion to future work.

Comparing the positive energy of mass and the negative energy of space leads to the prediction of a marginal distance $\bar \Omega$ beyond which gravitational action becomes repulsive. This distance compares well to the measured size of large voids in the universe \cite{Mueller2000}: massive objects at the rim of the voids repel each other such that they cannot enter the voids by gravitational forces.

Here, another fundamental problem arises: if energy is conserved, according to Noether's theorem, the universe must be stationary. We may recall the concept of self-similar distributions in materials science, e.g., coarsening in a multi-grain structure \cite{HILLERT1965,Darvishi2015}. Defining the relative length scale $\omega_r=\dfrac{\Omega}{\bar\Omega}$, one may reformulate the theory in this relative coordinate system. In the classical concepts cited above, the system reaches a self-similar distribution in the relative coordinates that is stationary in time: the system has to become self-similar if the system parameters are time-independent. We may assume that such a condition also holds for the universe, but a rigorous proof has not been given to date.


In the present treatment, space separates and connects particles. This defines the structure of the network of physical reality. Particles and spaces are two aspects of the doublons, the primitive object of the network; the doublon network of positive and negative states of energy defines the physical reality. It can be embedded into a 3D vector space or higher, but these spaces are not physical. In a multi-dimensional vector space, whether locally or globally defined, points are compact in any direction. In a network, points are compact only in one direction: the line coordinate $s$ between nodes. Points in a multi-dimensional vector space that do not lie on edges or nodes of the network---i.e., which do not coincide with at least one doublon---are inaccessible: they have no physical reality. 

A last comment shall be given regarding `general relativity.' General relativity is based on the premise that space is a multidimensional continuum which, in prevailing mathematical language, is formulated as a vector space. In this interpretation---the extroverted view of space---mass and space are treated as separate entities. Einstein's equations present a closure of the non-convex problem of space and mass: they connect masses via the deformation of space. The closure is based on Riemann curvature, which is basically derived from a network structure of discrete spaces. 

In the introverted view of space, where mass and space are part of the same entity (energy), the connection of mass and space emerges naturally. Energy is structured in the form of phases. Introducing space as a 1D line coordinate, one finds the doublon solution of phase expanded in this coordinate. The doublons are connected by the sum constraint \eqref{universe} to form a `universe.' Applying the gradient contribution of the Hamiltonian to the phases defines the positive energy of mass. Applying the gradient operators to the wave function and solving the quantum problem of acoustic excitation within the doublons defines the negative energy of the spaces. The theory defines gravitational attraction as a wave phenomenon. It is relativistic invariant and predicts repulsive gravitational action at ultra-long distances.

If we examine the arguments favoring the introverted view of space against the extroverted view, we are driven to conclude that multidimensional continuum space does not exist: it is a construct of our brains---an illusion.


\bibliography{./references}

@book{lectures,
  title={Lectures on Phase Field},
  author={Steinbach, I. AND Salama, H.},
  isbn={978-3-031-21170-6},
  url={https://link.springer.com/book/10.1007/978-3-031-21171-3},
  year={2023},
  publisher={Springer}
}

@article{Steinbach_QPF,
author={Steinbach, I.},
title={Quantum-phase-field concept of matter: Emergent gravity in the dynamic universe},
journal={Zeitschrift fur Naturforschung A},
year={2017},
volume={72},
number={1},
doi={10.1515/zna-2016-0270},
document_type={article},
}

@article{Steinbach_Soliton,
author={Kundin, J. and Steinbach, I.},
title={Quantum-phase-field: from the {B}roglie--{B}ohm double-solution program to doublon networks},
journal={Zeitschrift f\"{u}r Naturforschung},
year={2020},
volume={75},
number={2a},
doi={10.1515/zna-2019-0343},
document_type={article},
}

@Article{Steinbach2020,
  Title                    = {Erratum to: Quantum-phase-field concept of matter: Emergent gravity in the dynamic universe},
  Author                   = {I.~Steinbach},
  Journal                  = {Zeitschrift fur Naturforschung A},
  Year                     = {2020},
  Pages                    = {89-91}
}

@article{Steinbach99b,
author={Steinbach, I. and Pezzola, F.},
title={A generalized field method for multiphase transformations using interface fields},
journal={Physica D},
year={1999},
volume={134},
doi={10.1016/S0167-2789(99)00129-3},
document_type={article},
}

@article{DeWitt1967,
  Title                    = {Quantum theory of gravity. {I}. {T}he canonical theory},
  Author                   = {D. S. DeWitt},
  Journal                  = {Physical Review},
  Year                     = {1967},
  Pages                    = {1113--1148},
  Volume                   = {160}
}

@inproceedings{deBroglie1960,
  author = {L. de~Broglie},
  title = {Nonlinear wave mechanics},
  note = {Trans. A. J. Knodel, Elsevier (1960)},
  year = {1960},
}

@article{deBroglie1971,
  author = {L. de~Broglie},
  title = {L’interpretation de la mechanique ondulatoire par la theorie de la double solution},
  journal = {Proceedings of the International School of Physics Enrico Fermi},
  year = {1971},
  volume = {49},
  pages = {346--367}
}

@article{Bohm1952a,
  author = 	 {D. Bohm},
  title = 	 {A Suggested interpretation of the quantum theory in terms of ``hidden'' variables. I},
  journal =  {Physical Review},
  year = 	 {1952},
  volume = 	 {85},
  pages = 	 {166--179}
}

@article{Bohm1952b,
  author =  {D. Bohm},
  title =  {A suggested interpretation of the quantum theory in terms of ``hidden'' variables. II},
  journal =  {Physical Review},
  year =  {1952},
  volume =  {85},
  pages =  {180--193}
}

@article{Casimir1948,
  author = 	 {H. Casimir},
  title = 	 {On the attraction between two perfectly conducting plates},
  journal =  {Proceedings of the Koninklijke Nederlandse Akademie van Wetenschappen},
  year = 	 {1948},
  volume = 	 {B51},
  pages = 	 {793--795},
}

@article{Einstein1918,
  Title                    = {\"Uber {G}ravitationswellen},
  Author                   = {A. Einstein},
  Journal                  = {Sitzungsberichte der Königlich Preussischen Akademie der Wissenschaften Berlin},
  Year                     = {1918},
  Pages                    = {154--167},
  Volume                   = {}
}

@Article{Persinger2009,
author = {M.A.~Persinger},
title = {A Simple Estimate for the Mass of the Universe: Dimensionless Parameter {A} and the Construct of "Pressure"},
journal = {J. of Physics Astrophysics and Physical Cosmology},
year = {2009},
volume = {3},
pages = {1--3},
}

@ARTICLE{Goldstone1961,
       author = {{Goldstone}, J.},
        title = "{Field theories with  Superconductor solutions}",
      journal = {Il Nuovo Cimento},
         year = 1961,
        month = jan,
       volume = {19},
       number = {1},
        pages = {154-164},
          doi = {10.1007/BF02812722},
       adsurl = {https://ui.adsabs.harvard.edu/abs/1961NCim...19..154G}
}

@article{Goldstone1962,
  title = {Broken Symmetries},
  author = {Goldstone, J. and Salam, A. and Weinberg, S.},
  journal = {Phys. Rev.},
  volume = {127},
  issue = {3},
  pages = {965--970},
  numpages = {0},
  year = {1962},
  month = {Aug},
  publisher = {American Physical Society},
  doi = {10.1103/PhysRev.127.965},
  url = {https://link.aps.org/doi/10.1103/PhysRev.127.965}
}

@Article{Mueller2000,
  author = 	 {V.~M\"uller and S.~Arbabi-Bidgoli and J.~Einasto and D.~Tucker},
  title = 	 {Voids in the Las Campanas Redshift Survey versus cold dark matter models},
  journal =  {Mon. Not. R. Astron. Soc.},
  year = 	 {2000},
  volume = 	 {318},
  pages = 	 {280--288},
}

@article{HILLERT1965,
title = {On the theory of normal and abnormal grain growth},
journal = {Acta Metallurgica},
volume = {13},
number = {3},
pages = {227-238},
year = {1965},
issn = {0001-6160},
doi = {https://doi.org/10.1016/0001-6160(65)90200-2},
url = {https://www.sciencedirect.com/science/article/pii/0001616065902002},
author = {M Hillert},
}

@article{Darvishi2015,
author = {R. Darvishi Kamachali and A. Abbondandolo and K. F. Sieburg and I. Steinbach},
title = {Geometrical grounds of mean field solutions for normal grain growth},
journal = {Acta Materialia},
volume = {90},
OPTnumber = {},
OPTmonth = {01},
year = {2015},
OPTpages = {252258},
OPTnote = {},
OPTkey = {grain growth; grain size distribution; selfsimilar regime; mean field theory; phasefield simulation},
DOI = {10.1016/j.actamat.2015.02.025}
}

@article{Willox,
      title={de Broglie's double solution program: 90 years later}, 
      author={Samuel Colin and Thomas Durt and Ralph Willox},
      journal = {Annales de la Fondation Louis de Broglie},
      volume = {42},
      year = {2017},
      OPTpages = {19–71},
      DOI = {}
}

@article{Chiatti,
      title={Quantum Entities and the Nature of Time}, 
      author={Chiatti Leonardo},
      journal = {Qeios, https://doi.org/10.32388/5UTZO4}, 
      volume = {},
      year = {2023},
      OPTpages = {https://doi.org/10.32388/5UTZO4},
      DOI = {https://doi.org/10.32388/5UTZO4}
}

\bibliographystyle{unsrt}

\end{document}